\input amstex
\input xy
\xyoption{all}
\documentstyle{amsppt}
\document
\magnification=1200
\NoBlackBoxes
\nologo
\hoffset1.5cm
\voffset2cm
\def\em{\it}

\def\R{\bold{R}}

\def\rm{\roman}
\pageheight {16cm}
%\hfill{\it file Amywork/codes-complexity.tex, version 03.03.2012}

\bigskip

\centerline{\bf KOLMOGOROV COMPLEXITY AND THE ASYMPTOTIC BOUND}
%\smallskip
%\centerline{\bf } 
\medskip
\centerline{\bf  FOR ERROR--CORRECTING CODES}

\bigskip

\centerline{\bf Yuri I.~Manin${}^1$, Matilde Marcolli${}^2$}

\medskip

\centerline{\it ${}^1$Max--Planck--Institut f\"ur Mathematik, Bonn, Germany,}
\smallskip
\centerline{\it ${}^2$California Institute of Technology, Pasadena, USA}
\vskip1cm

\hfill{\it To the memory of Friedrich Hirzebruch,}

\hfill{\it a great mathematician and friend}

\bigskip

ABSTRACT. The set of all error--correcting block codes over a fixed
alphabet  with $q$ letters  determines a recursively enumerable set of rational points
in the unit square with coordinates $(R,\delta )$:= {\it (relative transmission rate,
relative minimal distance).} Limit points of this set form a closed subset,
defined by $R\le \alpha_q(\delta )$, where $\alpha_q(\delta )$
is a continuous decreasing function called {\it asymptotic bound.}
Its existence was proved by the first--named author in 1981 ([Man1]), but
no approaches to the computation of this function are known,
and in [Man5] it was even suggested that this function might be 
uncomputable in the sense of constructive analysis.

\smallskip

In this note we show  that the asymptotic bound becomes computable with the assistance of an oracle
producing codes in the order of their growing Kolmogorov complexity.
Moreover, a natural partition function involving complexity
allows us to interpret the asymptotic bound as a curve dividing two different
thermodynamic phases  of codes.

\vskip1cm

\centerline{\bf Introduction}

\medskip

In this article, we address again two related basic problems about
asymptotic bounds for codes, discussed  recently in [Man5]
and in [ManMar]. The first one is the problem of
{\it computability}, or, more suggestively, {\it plottability}
of the bound. The second one is the problem of interpretation
of this bound as a kind of phase--transition curve.

\smallskip

We start the sec.~1 below with precise definitions of the relevant notions,
and the reader may wish to turn to it immediately. Here we
restrict ourselves to some explanations on the intuitive level.  

\smallskip

Consider all error--correcting block codes $C$ in a fixed alphabet $A$ with $q$
letters. Each such code determines its {\it code point} $(R(C), \delta (C))$
in the plane {\it ($R$:= transmission rate, $\delta$:= minimal relative Hamming distance)}.

\smallskip
The {\it asymptotic bound $R=\alpha_q (\delta )$} is a continuous
curve in the plane  $(R,\delta )$ such that all limit points of the set of code
points lie below or on this bound, whereas all isolated code points lie above
it. 
\smallskip

Since 1981, when the existence of the asymptotic bound (and its versions for various structured codes,
such as linear ones) was discovered in [Man1],  many estimates for it from above and from below
were established, but no exact formula was found. This led one of us to conjecture in [Man5]
that the function $R=\alpha_q(\delta )$ might be {\it uncomputable} (and its graph unplottable)
in the technical sense formalized in [La] and [BratWe].

\smallskip

Here we treat codes assuming that {\it an oracle is given} that produces them in the order of {\it increasing
Kolmogorov complexity}, and show that with the assistance of such an oracle $R=\alpha_q(\delta )$
becomes ``plottable'' (sec.~2), and  that  appropriate partition functions involving 
either the Kolmogorov complexity, or Levin's prefix complexity ([Lev1])
change behavior across
 this asymptotic bound (sec.~3). (At the end of [Man3] it was argued that civilization is such a
universal oracle).

\smallskip

Slightly more precisely, it is known (see [CoGo], [BaFo], [VlaNoTsfa], and sec. 1 below) that if one 
chooses first a natural enumeration of codes and then generates 
error--correcting codes in the order of their ``size''
(actually, any computable order rather than complexity), then the density of the respective code points
is concentrated below or near the  bound (1.2) (for linear codes (1.3)), that lies 
in turn strictly below the asymptotic bound. 
Complexity is invoked here principally in order to identify typical random codes
with codes whose complexity is comparable with their size: cf. [ZvoLe], [Lev2], [LiVi]

\smallskip
By contrast, we show that code points of codes generated in the order of growing complexity,
that puts a considerable amount of highly non--random codes at the foreground, tend to be well 
distributed below the asymptotic bound,
with the bound itself appearing as a ``silver lining'' of the cloud of code points.
\smallskip

The last sec. 4 is dedicated to a sketch of ``quantization'' of the classical ensemble of codes.

\bigskip

\centerline{\bf 1. Asymptotic bound as a non--statistical phenomenon}

\medskip

{\bf 1.1. Codes and code points.} Here we recall our main definitions
and results of previous works.
\smallskip
We choose and fix an integer $q\ge 2$ and a finite set, {\it alphabet} $A$, or $A_q$,
of cardinality $q$. An (unstructured) {\it code} $C$ is defined as a non--empty subset $C\subset A^n$
of words of length $n\ge 1$.  Such $C$ 
determines its {\it code point} 
$$
P_C= (R(C),\delta (C))
$$ 
in the $(R,\delta )$--plane,  where $R(C)$ is
called  {\it the transmission rate}
and $\delta (C)$ is {\it the relative minimal distance of the code.}  They are defined by
the formulas
$$
\delta (C):= \frac{d(C)}{n(C)}, \quad  
d(C) := \roman{min}\,\{d(a,b)\,|\,a,b\in C, a\ne b\},\quad n(C):=n,
$$
$$
R(C) = \frac{[k(C)]}{n(C)}, \quad  k(C):=\roman{log}_q \roman{card}(C),
\eqno(1.1)
$$
where $d(a,b)$ is the Hamming distance
$$
d((a_i),(b_i)):= \roman{card} \{i\in (1,\dots ,n)\,|\,a_i\ne b_i\}.
$$
In the degenerate case $\roman{card}\,C=1$ we put $d(C)=0.$
We will call the numbers  $k=k(C)$, $n=n(C)$, $d=d(C)$, {\it code parameters}
and refer to $C$ as an $[n,k,d]_q$--code.  We denote by $Codes_q$ the set of all such codes,
and by $cp:\, Codes_q \to [0,1]^2\cap \bold{Q}^2$ the map $C\mapsto P_C$.
{\it The multiplicity} of a code point $x$ is defined as cardinality of the fiber $cp^{-1}(x)$.
\smallskip
If $q$ is a prime power, and $A_q$ is endowed with a structure of a finite field $\bold{F}_q$,
then {\it a linear code} is a linear subspace of $A_q^n$. The set of linear codes is denoted $Codes_q^{lin}$.
\smallskip

Our starting point here will be the following characterization of the set of all code points,
first proved in its final form in [Man5]. Note that $R$--axis is traditionally drawn as vertical one.

\medskip

{\bf 1.2. Theorem.} {\it There exists such a continuous function $\alpha_q (\delta)$, $\delta\in[0,1]$,
that 

\smallskip

(i) The set of code points of infinite multiplicity is exactly the set of  rational points  $(R,\delta )\in [0,1]^2$
satisfying $R\le \alpha_q(\delta )$. The curve  $R= \alpha_q(\delta )$ is called the asymptotic bound.
\smallskip

(ii) Code points $x$ of finite multiplicity all lie above  the asymptotic bound and are isolated: for
each such point  there is an open neighborhood  containing $x$ as the only code point.
\smallskip

(iii) The  same statements are true for linear codes, with, possibly, different
asymptotic bound $R= \alpha_q^{lin} (\delta )$.}

\medskip

{\bf 1.3. Good codes.}  One characteristic of a good code is this: {\it it maximizes 
simultaneously the transmission rate and the minimal distance.}  From this perspective, good
codes are isolated ones or lying close to the asymptotic bound. Below we briefly
describe known results showing that "most" randomly chosen  codes are  not good.
To the contrary, in the next section we show that in order to recognize good codes one must generate codes
of low Kolmogorov complexity, that is codes allowing short programs producing them.

\smallskip

This is exactly what has happened historically, when algebraic geometric codes, discovered by Goppa, were used 
by Tsfasman, Vladut and Zink in order to ameliorate the Gilbert--Varshamov bound: cf. an early
survey [ManVla] and [VlaNoTsfa].

\smallskip

In this sense, moral of this note is just opposite to the title of [CoGo]:
{\it Only codes about which we can think can be good.}

\medskip

{\bf 1.4. Shannon's ensemble.} We sketch here some well known arguments
and results
(see e.~g. [CoGo], [BaFo], [VlaNoTsfa] and references therein)
 showing that most (unstructured) $q$--ary codes lie lower or only slightly above the
curve
$$
R=\frac{1}{2}(1-H_q(\delta))
\eqno(1.2)
$$
 where $H_q(\delta )$ for $0<\delta<1$
is the $q$--ary entropy function
$$
H_q(\delta ) =\delta\roman{log}_q (q-1)-\delta \roman{log}_q \delta -
(1-\delta ) \roman{log}_q (1-\delta ).
$$
Notice that inside $[0,1]^2$ lies only the part of this curve for which $0\le R\le 1/2.$

\smallskip

The Gilbert--Varshamov bound 
$$
R=1-H_q(\delta)
\eqno(1.3)
$$ 
plays a similar role for linear codes: cf. Remarks below.

\smallskip

In order to make the statements above precise, one introduces Shannon's Random Code Ensemble 
$RCE_n$ of $q$--ary codes of block length $n$.  Each code in $RCE_n$ is a set of
pairwise different words in $A_q^n$ chosen randomly and independently with uniform
probability   $q^{-n}$.

\medskip

{\bf 1.5. Proposition ([CoGo, sec.~V]).} {\it (i) For any
$\varepsilon >0$,  the probability  (in $RCE_n$) that 
$H_q(d/n)\ge \roman{max} (1-2R,0)+\varepsilon$, where $R=k/n$, is bounded 
by $q^{-\varepsilon n(1+o(1))}$ as $n\to \infty$.
\smallskip

(ii) Similarly, the probability that $H_q(d/n)\ge 1-R+\varepsilon$ is bounded by
$e^{-q^{\varepsilon n(1+o(1))}}$ as $n\to \infty$.}

\medskip

{\bf Strategy of the  proof.} One easily sees that the number of words
at a (Hamming) distance $\le d$ from any fixed word in $A_q^n$ is
$$
{\rm Vol}_q(n,d) =\sum_{j=0}^d \binom{n}{j} (q-1)^j .
\eqno(1.4)
$$
As is well known, one can estimate this quantity in terms of the $q$--ary
entropy:
$$
q^{(H_q(\delta)-o(1))n} \leq {\rm Vol}_q(n, n\delta) \leq q^{H_q(\delta)n}.
$$
Following [CoGo], sec.~V, denote by $X^{(d)}$  the random variable on $RCE_n$ whose value at a code
is the number of  unordered pairs of distict code words at a distance $\le d$ from each other.
Clearly, on codes of cardinality $q^k$, we have from (1.4)
$$
E(X)=\binom{q^k}{2} \frac{\sum_{l=1}^d\binom{n}{l}(q-1)^l}{q^n-1}=q^{n[H_q(d/n)-(1-2R)] +o(n)}.
$$
One can similarly calculate $E(X^2)$, and to use Chebyshev's inequality to prove (i).
The last statement is obtained along the same lines. For details, see [CoGo].

\medskip

{\bf 1.6. Remarks.} For unstructured code points $(R,\delta)$ with $1-H_q(\delta) < 2R$ the same reasoning
shows so that the average number of pairs at distance $d$ is  large.
\smallskip

In the case of linear codes, the relevant code points concentrate in an exponentially narrow
neighborhood of the VG bound.

\smallskip

Since linear codes have considerably smaller Kolmogorov complexity than the general ones,
this behavior is compatible with our discussion in sec 2. below.

\medskip

{\bf 1.7. Spoiling operations as computable functions on codes.} The proof of existence 
of asymptotic bound (essentially, the only known one) is based upon
the existence of three types of rather banal combinatorial operations on (general, or linear)
codes  that produce from a given code several codes with worse parameters.
The subsequent combinatorial characterization of isolated code points 
as points of finite multiplicity, and proof  in [Man5] that any point with rational coordinates
below the asymptotic bound is a code point crucially uses these
operations as well.

\smallskip

In each class, the result of application of such an operation to a given code is by no means unique.
Here, using the discussion in [ManMar], sec. 1 (that reproduces
several earlier sources), we
will define three total recursive {\it spoiling maps} 
$$
S_i:\,Codes_q\to Codes_q, \quad i=1,2,3,
$$
that are compatible with the map $C\mapsto [n(C), k(C), d(C)]_q$ and 
whose effect on code parameters is summarized below:
$$
S_1:\ [n,k,d]_q \mapsto [n+1,k,d],
$$
$$
S_2:\ [n,k,d]_q \mapsto [n-1,k,d-1]\quad  (if\ n>1, k>0),
$$
$$
S_3:\ [n,k,d]_q \mapsto [n-1,k^{\prime},d],\quad where\ k-1\le k^{\prime}<k\quad  (if\ n>1, k>1).
$$
In order to make $S_i$ unambiguous, we simply choose the code with minimal number
wrt some computable numbering from
a finite set of codes 
that can be obtained in this way, and define it to be $S_i(C)$.
\smallskip
Another (minor) point is to decide what to do if mild restrictions in
brackets do not hold for $C$. The simplest solution that we adopt
is to put then $S_i(C)=C.$ 
\medskip

{\bf 1.8. Block length and distance between isolated codes.} We will show now that
knowing the distance of an isolated code point $x$ from its closest neighboring code point,
we can estimate from above the maximal block length of a code mapping to $x$
and hence also the multiplicity of $x$. Distance in $[0,1]^2$ is defined here
as $\roman{dist} ((a,b),(c,d)):=\roman{max}(|a-c|,|b-d|)$.

\medskip

{\bf 1.8.1. Proposition.} {\it Let $(R,\delta)$ be an
isolated code point  and denote by $\rho$ its distance
from the closest neighboring code point. Assume that $(R,\delta)$ is the code point
of some $C\subset A^N.$ In this case we have
$$
N\le \roman{max}\left( \frac{R-\rho}{\rho}, \frac{\delta-\rho}{\rho}\right)
$$ 
}
\medskip

{\bf Proof.}  If the code parameters of  $C$ are $[N,K,D]_q$,
then the code parameters of $S_1(C)$ are $[N+1,K,D]_q$ so that
$$
\roman{dist} (P_C, P_{S_1(C)})= \roman{max} \left(\frac{[K]}{N+1}, \frac{D}{N+1}\right)\ge \rho.
$$
This shows our result, because 
$$
R=\frac{[K]}{N},\ \delta= \frac{D}{N}.
$$

\bigskip

\centerline{\bf 2. Plotting asymptotic bound with assistance of a complexity oracle}

\medskip

{\bf 2.1. A general setup.}  Let $X$ be an infinite constructive world,
in the sense of [Man3]. This means that we have
a set of {\it structural numberings} of $X$: computable  bijections $\bold{Z}^+\to X$,
forming a principal homogeneous space over the group
of total recursive permutations $\bold{Z}^+\to \bold{Z}^+$
\smallskip
 Consider one such  bijection $\nu=\nu_X:\,  \bold{Z}^+\to X$.
It defines an order on $X$: $ x^{\prime}\le x$ iff $\nu^{-1}( x^{\prime})\le \nu^{-1} (x)$.
 Equivalently,  we  can imagine such a bijection as an order in which
elements of $X$ are generated: $x$ is generated
at the $\nu(x)$--th step. 
\smallskip
Another important class of bijections that we have in mind consists of (uncomputable) {\it Kolomogorov
orders} defined and discussed in [Man3]. Namely, let $u:\,\bold{Z}^+\to X$
be an optimal  (in the sense of Kolmogorov and Schnorr) partial recursive enumeration. Then
$K_u(x):= \roman{min}\,\{k\,|\,u(k)=x\}$ is the (exponential) Kolmogorov complexity, 
and the Kolmogorov order of $X$
is the order of growing complexity.  If we denote the respective Kolmogorov order
$\bold{K}_u(x)$, we have $c_1K_u(x)\le  \bold{K}_u(x) \le  c_2K_u(x)$ for constants $c_1,c_2>0$
depending only on $u$.  Similarly, another choice of $u$ produces another
order  differing from $\bold{K}_u$ by a permutation of linear growth.

\smallskip

Let now $X,Y$ be two infinite constructive worlds, $\nu_X, \nu_Y$ respective structural bijections, 
$u:\,\bold{Z}^+\to X $,  $v:\ \bold{Z}^+\to Y$ two optimal enumerations, and   $K_u, K_v$ 
the respective Kolmogorov 
complexities.
\smallskip
Consider a total recursive function $f:\,X\to Y$. 

\medskip

{\bf 2.2. Proposition.} {\it For each $y\in f(X)$, there exists $x\in X$ such that  
$$
y=f(x), \quad K_u(x)\le \roman{const}\cdot \nu_Y^{-1}(y)
\eqno(2.1)
$$
where the constant can be calculated in terms of $u,v,\nu_X, \nu_Y.$
}
\medskip
{\bf Proof.} Informally, this means that we can effectively generate all elements of 
the (enumerable) image $f(X)\subset Y$ in their structural order,  
 if an oracle generates for us all elements of $X$ in the order of growing  Kolmogorov
 complexity.
 
 \smallskip
 
 In fact, denote by $F:\, X\to Y\times \bold{Z}^+$ the following 
map:
$$
F(x):= (f(x),n(x)),\quad \roman{where}\ n(x):=\roman{card}\,\{x^{\prime}\,|\, \nu_{X}^{-1}(x^{\prime})\le {\nu}_X ^{-1}(x),f(x^{\prime})=f(x)\}.
$$
In plain words, $n(x)$ is the number of $x$ in the set $f^{-1}(f(x))$ wrt the order induced by $\nu_X$.
\smallskip
Clearly,
$F$ is a (total) recursive function.  
Hence the image $E\subset Y\times \bold{Z}^+$  of $F$ 
is an enumerable subset of  $Y\times \bold{Z}^+$.
\smallskip
For each $m\in \bold{Z}^+$, put
$$
X_m:= \{ x\in X\,|\, n(x)=m \} \subset X,\quad Y_m:=f(X_m)\subset Y.
$$
Then $X_m$ (resp. $Y_m$) is an enumerable subset of  $X$ (resp. $Y$),
and  the restriction of $f$ upon $X_m$ induces a  bijection of these sets.
Moreover, $f(X_1)=f(X)$, so that we can define the partial recursive function 
$\varphi :\, Y\to X$, with domain $f(X)$, which is $f^{-1}$ on its domain.

\smallskip

Hence, for any $x\in X_1$  such that $f(x)=y$,
we will have
$$
K_u(x)=K_u(\varphi (y))    \le c_{\varphi}\, K_v(y)\le c^{\prime}\, \nu_Y^{-1}(y).
$$
for appropriate constants $c_{\varphi}, c^{\prime}$. 
Here and below we use some basic inequalities involving
complexities, proved e.g. in [Man4], VI.9.
\smallskip
This completes the proof.

\medskip

{\bf 2.3. Finite vs infinite multiplicity.} We keep notation of the previous subsections,
in particular, $f:\,X\to Y$ is a total recursive function. For any $y\in Y$, call
$$
\roman{mult}\,(y):=\roman{card}\,f^{-1}(y)
$$ 
{\it the multiplicity} of $y$. The Proposition 2.2 shows
that with assistance of a complexity oracle for $X$ we can generate consecutively 
elements of $Y$ of nonzero multiplicity. For applications to codes, we want
to divide them into two basic subsets: elements of finite multiplicity $Y_{fin}$
(they will correspond to isolated code points) and  elements of infinite multiplicity $Y_{\infty}$.
The latter will correspond to code points lying below or on the asymptotic bound.  

\smallskip

From the definitions above, it is clear that 
$$
Y_{\infty} \subset \dots \subset f(X_{m+1})\subset f(X_m)\subset \dots \subset f(X_1)=f(X),
$$ 
and
$$
Y_{\infty} = \cap_{m=1}^{\infty} f(X_m),\quad Y_{fin}= f(X)\setminus Y_{\infty} .
$$

We have the following extension of the Proposition 2.2.

\medskip

{\bf 2.3.1.  Proposition.} {\it For each $y\in Y_{\infty}$ and each $m\ge 1$, there exists unique $x_m\in X$ such that  
$y=f(x_m)$, $n(x_m)=m$, and
$$
 \quad K_u(x_m)\le \roman{const}\cdot \nu_Y^{-1}(y)m\,\roman{log}(\nu_Y^{-1}(y)m)
 \eqno(2.2)
 $$
where the constant does not depend on $y,m$ and can be calculated in terms of $u,v,\nu_X, \nu_Y$.
}
\medskip
{\bf Proof.} Define the partial function
$$
\Phi:\, Y\times \bold{Z}^+ \to X
$$
with domain
$$
D(\Phi ):= \{\,(y,m)\,|\,\roman{mult}\,(y)\ge m\,\}
$$
such that 
$$
\Phi(y,m) := \roman{the\  }m-\roman{th\ element\ in\ the\ fiber\  }f^{-1}(y).
$$ 
One easily sees that its graph is enumerable, hence $\Phi$ is partial recursive.
The element $x_m$ in the statement of Proposition is just $\Phi (y,m)$. 
Therefore
$$
K_u(x_m)= K_u(\Phi (y,m))\le \roman{const}\cdot K((\nu_Y^{-1}(y),m)).
$$
One can get various estimates of a chosen Kolmogorov complexity $K$ of the pair $(\nu_Y(y),m)$
by choosing various structural  numberings of the product of two
constructive worlds $Y\times \bold{Z}^+$:  cf. a more detailed discussion in
sec. 2.7--2.10 of  [Man4]. Here we use the standard estimate symmetric in both
arguments: 
$$
K((\nu_Y(y),m))\le \roman{const}\cdot K(\nu_Y(y)) K(m)\, \roman{log} (K(\nu_Y(y)) K(m)).
$$
This completes the proof, since complexity on $\bold{Z}^+$ is majorized by identical function.

\medskip

Now, an oracle mediated process of generating sets $Y_{\infty}$, $Y_{fin}$, 
can be described as the following
inductive procedure. Choose a sequence of pairs of positive integers $(N_m,m)$, $m=1,2,3 \dots$, 
$N_{m+1}>N_m$.
\medskip

{\it Step 1.} Produce the list of
all elements $y\in f(X)$ such that $\nu_Y^{-1}(y)\le N_1.$ Denote this list $A_1$, and put $B_1=\emptyset$.

\smallskip

If lists (subsets)  $A_m,B_m \subset f(X)$ are already constructed at the $m$--th step, go to

\smallskip

{\it Step m+1.}  Produce the list of
all elements $y\in f(X)$ such that $\nu_Y^{-1}(y)\le N_{m+1}.$ Denote by
$A_{m+1}$ the subset of  elements $y$ in this list for which there exists $x\in X$
with $f(x)=y$, $n(x)=m+1$, and denote by $B_{m+1}$ the subset of remaining elements.
According to (2.2), we will have to ask the oracle to produce the list of  $x\in X$
with explicitly bounded complexity, in order to be sure that this $x$ with $n(x)=m+1$
appears in this list, if it exists at all.

\smallskip

It is clear that $A_m\cup B_m\subset  A_{m+1}\cup B_{m+1}$, and that the union
of this sequence of sets is $f(Y)$. Moreover,
$B_m\subset B_{m+1}$ and $\cup_{m=1}^{\infty}B_m= Y_{fin}$.
The passage from $A_m$ to $A_{m+1}$ generally involves both adding new elements $y$
(with $N_m<\nu_Y^{-1}(y)\le N_{m+1}$) and forwarding some of the elements of $A_m$  to $B_{m+1}$
(whenever it turns out that in their fiber no new element of $X$ appears).

\medskip

{\bf 2.4. A structural numbering of $q$--ary codes.} We will now explain how the
general procedure described above can be applied to codes. More precisely,
we will describe the constructive world of $q$--ary codes  $X=Codes_q$,
the constructive world of  rational points $Y=[0,1]^2\cap \bold{Q}^2$ and the
total recursive map $f:X\to Y,\, C\mapsto cp(C)$.

\medskip

 We fix $q$ and the alphabet $A$ of cardinality $q$.
We fix also a total order on $A$, say, by identifying $A$ with $\{0,1,\dots , q-1\}.$
We then order all words in $A^n$ lexicographically.

\smallskip

Define now the following computable total order (or, equivalently, computable
bijective enumeration $\nu$) of the set of all non--empty codes $Codes_q$ with $k>0$:

\smallskip

a) If $n_1< n_2$, all codes in $A^{n_1}$ come before those in $A^{n_2}$.

\smallskip

b) If $k_1< k_2$, all  $[n,k_1,d]_q$--codes  come before  $[n,k_1,d^{\prime}]_q$--codes.

\smallskip

c) When $(n,q^k)$ are fixed,  order all words in $A^n$ lexicographically,
then consider induced order on words in any code $C\subset A^n$, and finally
encode $C$ by the concatenation of all its elements put in the lexicographic
order. Such a word $w(C)\in A^{nq^k}$  determines $C$ uniquely.
Finally, order all $[n,k,d]_q$--codes in the lexicographic
order of $w(C)$.

\smallskip

Clearly, $n(C), [k(C)]+1$ and  $d(C)$ become  total recursive functions $Codes_q\to \bold{Z}_+$
(condition $k(C)>0$ means that $C$ contains at least two different words the distance between which
is therefore positive).

Finally, the function ``code point'' 
$$
cp (C):= \left( \dfrac{[k]}{n},\dfrac{d}{n}\right)
$$
is a total recursive map $Codes_q\to [0,1]^2\cap \bold{Q}^2$.

\medskip

{\bf 2.5. Plotting asymptotic bound.} Now apply to the codes the general procedure discussed above.
Fix an enumeration $\nu_Y$ of rational points in the unit square in some natural way.
To make the picture visually transparent, choose the sequence $N_m$ in such a way
that  the set  of points with $\nu_Y^{-1}(y)\le N_m$ contains the subset $C_m$ of
 all points with denominators of each coordinate 
dividing $m!$, and plot at the step $m$ only the points of $A_m,B_m$ contained in $C_m$.
\smallskip

Clearly, ``pixels''  in $C_m$ form the vertices of the square lattice of radius $1/m!$.
Call a subset of $C_m$ {\it saturated}, if it is a union of sets of the form $S_{a,b}$:
$\{(x,y)\,|\,x\le a, y\le b\}$, $(a,b)\in C_m$.  To motivate this definition, recall that
the subset of $C_m$ lying under or on the asymptotic bound is saturated.
Hence it must be contained in the maximal saturated subset $D_m$ of $A_m\cap C_m$.

\smallskip

The upper boundary $\Gamma_m$ of this subset (consisting of horizontal and vertical segments of
the length $1/m!$ that connect neighboring points) is our $m$--th approximation to the asymptotic bound.

\smallskip

Obviously, $B_m$ is the subset of isolated codes constructed at the $m$--th step.

\smallskip

The status of  any point that is above $\Gamma_m$ but not contained in $B_m$
will become clear only at a subsequent step. 

\bigskip

\centerline{\bf 3. Partition functions for codes and phase transition effects}

\medskip

{\bf 3.1. Partition functions involving complexity.} Let $X$ be an infinite constructive world.
Following L.~Levin ([Lev1], [Lev2]), we will consider functions
$$
p:\, X\to \bold{R}_{>0}\cup \{\infty\},  
$$
that are {\it enumerable from below} in the sense that the set
$$
\{ (r, p(x))\,|\, r<p(x)\}    \subset \bold{Q}\times X
$$
is enumerable.

\smallskip

Furthermore, Levin introduces the notion of a {\it quasinorm functional} on the set
of enumerable from below functions and shows that for any choice of such a functional
$N$,  the set of functions $p$ with $N(p)<\infty$ admits a maximal one in the sense that
it majorizes any other function after multiplication by an appropriate positive constant.
\smallskip

We quote here two representative examples showing relation of this result to complexity:
\medskip

{\bf 3.1.1. Proposition.} {\it  (i) Let 
$$
N(p):=\roman{sup}\,(r\cdot \roman{card}\,\,\{x\,|\, p(x)\ge r\} ).
$$
For this quasinorm, functional $p(x):=K_u(x)^{-1}$ is a maximal function, where $K_u$ is a
Kolmogorov complexity on $X$.

\medskip

  (ii) Let 
$$
N(p):= \sum_{x\in X} p(x).
$$
For this quasinorm,  functional, $p(x):=KP_v(x)^{-1}$ is a maximal function, where $KP_v$ is an
(exponential) prefix--free complexity on $X$ depending on the respective optimal  ``decompressor'' $v$.
}

\medskip

Initial definition of prefix--free complexity involved the choice of the world of binary
words for $X$. However, Levin's result 3.1.1(ii)  gives a very natural independent
characterization of this complexity in terms of {\it enumerable from below probability distributions} on $X$
whose definition uses only the fact that $X$ is a constructive world.

\smallskip

This construction shows that it is natural to consider at least two types
of partition functions on a constructive world $X$ that endow objects of low complexity
with higher weight:  $Z(X,\beta )= \sum_{x\in X} K_u(x)^{-\beta}$ and
$ZP(X,\beta )= \sum_{x\in X} KP_v(x)^{-\beta}$ where $\beta$ is the inverse temperature.
Both absolutely converge for $\beta >1$ and diverge for $\beta <1.$
The difference is that the first one  diverges at $\beta =1$,
whereas the second one converges for $\beta =1$ as well. 
Divergence is easily seen, if one replaces Kolmogorov complexity
by Kolmogorov order, in which case the partition function
becomes simply Riemann's $\zeta (\beta )$.

\smallskip

In the following we will use versions of $Z(X,\beta )$, in particular, because
the usual Kolomogorov complexity was extended to the nonconstructive world
of partial recursive functions (e.~g. in [Sch] and in the first 1977 edition of [Man4]), and this allowed
 one to prove for it a number of useful estimates. Here is the simplest one.
 
 \medskip
 
{\bf 3.2. Lemma.} {\it Let  $Y$ be an infinite decidable subset of a constructive
world $X$. Endowing $Y$ with the induced structure of constructive world,
choose  two exponential Kolmogorov complexities $K_u(X,*)$, resp. $K_{v}(Y, *)$
of the objects in $X$, resp. in $Y$. Then the restriction of  $K_u(X,*)$ to $Y$
is equivalent to  $K_{v}(Y, *)$ in the sense that one of these functions multiplied by a positive
constant majorizes another one.}

\medskip

{\bf Proof.} The embedding $i:\, Y\to X$ is total recursive function.
Define the function $j:\, X\to Y$ as identity on $Y$, and taking a constant value
$y_0\in Y$ on the complement $X\setminus Y$. Since this complement
is enumerable, $j$ is total recursive as well. Hence
$K_u(i(y))\le c_1K_v(y)$, $K_v(j(x))\le c_2K_u(x).$

\medskip

{\bf 3.3. Phase transition.} Since the function $\alpha_q(\delta )$ is continuous
and strictly decreasing for $\delta \in [1,1-q^{-1})$, the limit points domain $R\le  \alpha_q(\delta )$
can be equally well described by  the inequality $\delta\le \beta_q(R)$
where $\beta_q$ is the function inverse to $\alpha_q$.

\smallskip

Fix now an $R\in \bold{Q} \cap (0,1)$. For $\Delta \in  \bold{Q} \cap (0,1)$, put
$$
Z(R, \Delta;\beta ):=   \sum_{C:\, R(C)=R,\Delta\le \delta (C) \le 1} K_u(C)^{-\beta +\delta (C) -1},
\eqno(3.1)
$$
where $K_u$ is an exponential Kolmogorov complexity on $Codes_q$.

\medskip

{\bf 3.3.1. Theorem.}  {\it  (i) If $\Delta > \beta_q(R)$, then $Z(R, \Delta;\beta )$ is a real analytic function of $\beta$.

\smallskip

(ii) If $\Delta < \beta_q(R)$,  then $Z(R, \Delta;\beta )$ is a real analytic function of $\beta$ 
for $\beta >\beta_q(R)$ such that its limit for $\beta -\beta_q(R)\to +0$ does not exist.
} 
\medskip

{\bf Proof.}  If $\Delta > \beta_q(R)$, then all codes in the summation domain of (3.1)
are isolated ones, and there is only a finite number of them, hence  $Z(R, \Delta;\beta )$
is real analytic. Otherwise, this set of codes is infinite decidable subset of $Codes_q$,
and one can appeal to 3.2.

\medskip

{\bf 3.3.2. Comments.} To embed the statement of Theorem 3.3.1 in a  conventional 
environment of thermodynamics, one should have in mind
the following analogies. The argument $\beta$ in (3.1) is the inverse temperature,
the transmission rate $R$ is a version of density $\rho$, so that our asymptotic bound
transported into $(T=\beta^{-1},R)$--plane as  $T=\beta_q(R)^{-1}$ becomes
the phase transition boundary in the (temperature, density)--plane.

\medskip

{\bf 3.4. Measures and the asymptotic bound.} We now show that the plotting
procedure described in Section 2 can be reformulated in terms of measures
on the space of codes defined by the partition functions described above. 

\smallskip

The partition function $Z(X,\beta)=\sum_{x\in X} K_u(C)^{-\beta}$ determines a 
one-parameter family of probability measures on the space $X$ of codes, 
for $\beta >1$, given by
$$ \bold{P}_\beta(C) = \frac{ K_u(C)^{-\beta} }{Z(\beta)}. $$
Similarly, one obtains probability measures associated to the partition
functions $ZP(X,\beta)$ and $Z(R,\Delta;\beta)$, with the latter defined on
the space of codes with parameter $R(C)=R$ and $1-\Delta\leq \delta(C)\leq 1$.

\smallskip

Now consider again the oracle mediated process described in Section 2,
generating the sets $Y_\infty=V_q\cap U_q$ and 
$Y_{fin}=V_q \smallsetminus (V_q \cap U_q)$ of code points below and
above the asymptotic bound, and the inductively constructed sets $A_m$ and $B_m$.

\medskip

{\bf 3.5. Proposition.} {\it The algorithm of Section 2 determines a sequence of 
probability measures associated to the sets $A_m$ and $B_m$ that converge 
to probability measures on the space of codes with parameters in $Y_{fin}$ 
and $Y_{\infty}$ and a sequence of measures $\bold{P}_{m,\beta}$ converging 
to a measure supported on the asymptotic bound curve.}

\medskip

{\bf Proof.} We work with the partition function $Z(X,\beta)$. The argument for
$ZP(X,\beta)$ is analogous.  On each of the sets $B_m$ constructed by the oracle
mediated algorithm of Section 2, one obtains an induced probability measure
$\bold{P}_{B_m,\beta}(C)=  K_u(C)^{-\beta} Z(cp^{-1}(B_m),\beta)^{-1}$,
where  $Z(cp^{-1}(B_m),\beta) = \sum_{C\in cp^{-1}(B_m)} K_u(C)^{-\beta}$.
Since all the code points in $Y_{fin}$ have finite multiplicity, and each $B_m$
contains finitely many code points, the $Z_{B_m}(\beta)$ are finite sums for 
all $m\geq 1$. Since the sets $B_m\subset B_{m+1}$ are nested, in the limit 
$m\to \infty$, the probability measures $\bold{P}_{B_m,\beta}(C)$ converge
to the probability measure supported on $cp^{-1}(Y_{fin})$ given by
$\bold{P}_{fin,\beta}(C) = K_u(C)^{-\beta} Z(cp^{-1}(Y_{fin}),\beta)^{-1}$
with $Z(cp^{-1}(Y_{fin}),\beta) =\sum_{C\in cp^{-1}(Y_{fin})} K_u(C)^{-\beta}$.

In the case of the sets $A_m$, one has $A_m= (A_m\cap A_{m+1}) \cup (A_m\cap B_{m+1})$,
and one obtains the set $Y_\infty=V_q\cap U_q$ of code points below the
asymptotic bound as 
 $$ Y_\infty = \bigcup_{m\geq 1} (\bigcap_{n\geq 0} A_{m+n}). $$
Consider the sequence of sets $E_{M,N}=\cup_{m=1}^M\cap_{n=0}^N A_{m+n}$.
Then $E_{M+1,N}\supset E_{M,N}$ and $E_{M,N+1}\subset E_{M,N}$.   
Correspondingly, one has sequences of probability measures 
$$ \bold{P}_{E_{M,N}} (C) = \frac{K_u(C)^{-\beta} }{ Z(cp^{-1}(E_{M,N}),\beta)}, $$
with $Z(cp^{-1}(E_{M,N}),\beta)=\sum_{C\in cp^{-1}(E_{M,N})} K_u(C)^{-\beta}$,
that converge, as $M, N\to \infty$ to the probability measure
$\bold{P}_{\infty,\beta}(C) = K_u(C)^{-\beta} Z(cp^{-1}(Y_{\infty}),\beta)^{-1}$,
supported on codes in $cp^{-1}(Y_{\infty})$ with 
$Z(cp^{-1}(Y_\infty),\beta) =\sum_{C\in cp^{-1}(Y_{\infty})} K_u(C)^{-\beta}$.

Consider then the sets $C_m$ and $D_m \subset A_m\cap C_m$
constructed as in Section 2, and the upper boundary $\Gamma_m$
of the set $D_m$ approximating the asymptotic bound. Denote by
$F_m \subset D_m$ the region bounded between $\Gamma_m$
and $D_m \cap \Gamma_{m-1}$. Then the probability measures
$$ \bold{P}_{F_m}(C) =\frac{  K_u(C)^{-\beta}}{ Z(cp^{-1}(F_m),\beta)}, $$
with $Z(cp^{-1}(F_m),\beta)=\sum_{C\in cp^{-1}(F_m)} K_u(C)^{-\beta}$
converge to a probability measure 
$\bold{P}_{\Gamma}(C) = K_u(C)^{-\beta} Z(\Gamma,\beta)^{-1}$, 
supported on the set of codes whose code points fall on the asymptotic bound 
curve $\Gamma=\{ (R,\delta)\,|\, R=\alpha_q(\delta) \}$ itself,
with $Z(\Gamma,\beta)=\sum_{C\in cp^{-1}(\Gamma)} K_u(C)^{-\beta}$.

\medskip

When using the partition function $Z(R,\Delta;\beta)$ of (3.1), one has an analogous
statement, with code points restricted to the domain $R(C)=R$ and 
$1-\Delta\leq \delta(C)\leq 1$, except for the last statement about a measure
supported at the asymptotic bound, because of the presence of a phase transition
for $Z(R,\Delta;\beta)$ precisely along that curve.

\bigskip

\centerline{\bf 4. From classical to quantum systems}

\medskip

{\bf 4.1. Computable functions as classical observables.} In the subsections 3.4--3.5
 we have described a statistical mechanical
system on the space of codes, where observables are
computable functions on the space $X$ of $q$--ary codes codes and the
expectation values of observables are obtained by
integrating these functions against the probability measure
defined by the complexities,
$$ \langle f \rangle_{\beta} = \int f(C) \, d\bold{P}_\beta(C) =
\frac{1}{Z(X,\beta)} \sum_{C \in X} f(C) \, K_u(C)^{-\beta}, 
$$
or similarly with the measures defined by $ZP(X,\beta)$ or
$Z(R,\Delta,\beta)$.

\medskip

In this section we describe a quantized version of this
statistical system and explain the role of the asymptotic
bound curve $R=\alpha_q(\beta)$ in this setting.

\smallskip

The quantization of the system is achieved by considering
code words as the possible independent degrees of freedom
in an unstructured code, and quantizing them as independent
harmonic oscillators, with energy levels that depend on the
rate and the complexity of the code. 

\smallskip

We then show that, while
code points that lie below the asymptotic bound give rise in
this way to a bosonic field theory with infinitely many degrees
of freedom, the code points above the asymptotic bound
produce quantum mechanical systems with finitely many
degrees of freedom. 

\smallskip

The partition function of the resulting
quantum statistical mechanical system is different from
the one we computed in Section 3 for the classical system,
but it is easily derived from it and displays similar phase
transition phenomena. 

\smallskip

We also show that the recursive algorithm of
Section 2 provides a good approximation by systems 
with finitely many degrees of freedom for the quantum
system associated to the set $Y_\infty$ of code parameters.

\medskip

{\bf 4.2. Quantum statistical mechanical system of a single code.}
To make a single unstructured code $C$ into a quantum system,
we regard the code words as the possible independent degrees 
of freedom and we associate to each of them a 
creation and annihilation operator. This means that we consider,
for each code word $x\in C$ an isometry $T_x$, with  $T_x^* T_x=1$
and such that the $T_x T_x^*$ are mutually orthogonal projectors.
This means that we associate to a given code the Toeplitz algebra 
${\Cal T}_C$ on its set of code words. This is the same kind of
code algebra as we considered in our previous work [ManMar].

\smallskip

The algebra ${\Cal T}_C$ is naturally represented on the
corresponding Fock space, namely the Hilbert space ${\Cal H}_C$ 
spanned by the orthonormal basis $\epsilon_w$ with $w=x_1\ldots x_N$ 
ranging over the set of finite sequences (of arbitrary length $N$)
of the code words $x\in C$. In this representation, the operator $T_x$
acts by appending $x$ as a prefix to a given string of code words, 
$T_x \epsilon_w= \epsilon_{xw}$.

\smallskip

The dynamics of this quantum system is determined by a Hamiltonian
operator $H$ on the Fock space, via the time evolution
$$ 
T \mapsto q^{it\,H} \, T\, q^{-it\, H}. 
$$
We can then assign energy levels that depend on the complexity of the
code in the following way.

\smallskip

{\bf 4.3. Lemma.} {\it The time evolution $\sigma: \R \to Aut({\Cal T}_C)$
given by $\sigma_t (T_x)= K_u(C)^{it}\, T_x$ is generated by the Hamiltonian
$H \epsilon_w = \ell(w) \log_q K_u(C)\, \epsilon_w$, with $\ell(w)$ the
length of the word $w$, and has partition function
$$ 
Z(C,\sigma,\beta) = \frac{1}{1-q^{nR}K_u(C)^{-\beta}}, \eqno(4.1) 
$$
which is a real analytic function in the domain
$\beta> nR/\log_q K_u(C)$. }

\medskip

{\bf Proof.} 
The Hamiltonian implementing the time evolution $\sigma_t (T_x)= K_u(C)^{it}\, T_x$
on the Fock space ${\Cal H}_C$ is the operator $H$ on ${\Cal H}_C$ satisfying
$$
 \sigma_t(A) = q^{it H} A \, q^{-it H},  \ \ \  \forall A\in {\Cal T}_C. 
 $$
This is given by the operator $H \epsilon_w = m \log_q K_u(C)\, \epsilon_w$ for
all words $w=x_1\ldots x_m$ of length $\ell(w)=m$. We then find 
$$ 
Z(C,\sigma,\beta)= Tr(q^{-\beta H})= \sum_m (\rm{card}\, W_m)  q^{-\beta m\, \log_q K_u(C)} =
\sum_m q^{m \, (nR -\beta \log_q K_u(C))} , 
$$
where the cardinality of the set $W_m$ of words of length $m$ is $q^{mk}$,
since $\rm{card}\, C =q^k=q^{nR}$, with $n(C)=n$ the length of the code and $R(C)=R$ the rate.
This series converges to (4.1) for $\beta > nR/\log_q K_u(C)$.

\medskip

To compare the behavior of this partition function to the $Z(R,\Delta;\beta)$
considered in Theorem 3.3.1, it is convenient to change variable in (4.1) by
$\beta \mapsto n(C) (\beta -\delta(C)+1)$. Then, using the Singleton
bound on codes, we obtain the following. 

\smallskip

{\bf 4.4. Corollary.} {\it The function $Z(C,\sigma,\alpha)$, with
$\alpha = n (\beta -\delta+1)$ is real holomorphic for all  $\beta>0$
with 
$$ 
Z(C,\sigma,n (\beta -\delta+1)) \leq  \frac{1}{1-K_u(C)^{-\beta}}. 
$$
}

\medskip

{\bf Proof.} The partition function $Z(C,\sigma,\alpha)$ is given by the sum
$$ 
\sum_m q^{mn (R -(\beta -\delta +1) \log_q K_u(C))} \leq 
\sum_m q^{m n( R+\delta -1 )} K_u(C)^{-\beta m} \leq  
\sum_m K_u(C)^{-\beta m}. 
$$
where the first estimate uses $K_u(C)^{\delta-1}\leq q^{\delta-1}$ and the
second estimate uses the singleton bound for codes $k \leq n-d-1$, which gives
$R+\delta -1\leq 0$. 

\medskip

{\bf 4.5. Quantum statistical mechanical system at a fixed code point.}
We can now consider quantum statistical mechanical systems involving
several codes. Again, the main idea is that different codes with their 
degrees of freedom given by their code words are considered as
independent uncoupled systems, which means that the algebra
of observables describing the set $X_{(R,\delta )}$ of $q$--ary codes with fixed code
parameters $(R,\delta)$ becomes the tensor product of
the Toeplitz algebras of the individual codes,
$$ 
{\Cal T}_{(R,\delta)}=\otimes_{C \in X_{(R,\delta )}} {\Cal T}_C
 \eqno(4.2) 
$$
acting on the tensor product of the Fock spaces
${\Cal H}_{(R,\delta)}=\otimes_{C \in X_{(R,\delta)}} {\Cal H}_C$ and
with the product time evolution $\sigma_t^{(R,\delta)}= \otimes_C \sigma_t^C$,
with $\sigma_t^C(T_x)=K_u(C)^{it} T_x$.
The partition function becomes, correspondingly, the product of the
partition function for the independent systems. In particular,
for $\alpha=\alpha(n,\beta,\delta)=n(\beta-\delta +1)$ a variable
inverse temperature as in [ManMar], we find that
$$
Z(X_{(R,\delta)},\sigma, \alpha) =\prod_{C \in X_{(R,\delta)}}
Z(C,\sigma,n(\beta-\delta +1)),
$$
is a finite product for $(R,\delta) \in Y_{fin}$ and an infinite product
for $(R,\delta) \in Y_\infty$, whose convergence is controlled by the convergence of
the infinite product
$$\prod_{C \in X_{(R,\delta)}} (1-K_u(C)^{-\beta})^{-1} .
$$ 
This 
in turn converges whenever  the series
$Z(X_{R,\delta},\beta)=\sum_{C \in X_{(R,\delta)}} K_u(C)^{-\beta}$ converges, which is the classical
partition function for a fixed code point.

\smallskip

This argument shows the role of the asymptotic bound from the point of view of
these quantized systems. In fact, recall that an infinite tensor
product of Toeplitz algebras is a standard way to describe the second
quantization of a bosonic field theory, see for instance [Jul] and also
Section 2 of [BoCo]. However, a finite tensor product is a purely quantum
mechanical system, with only finitely many degrees of freedom. Thus,
the asymptotic bound separates the region $Y_{fin}$ in the space of code
parameters where the quantum statistical system $({\Cal T}_{(R,\delta)},\sigma_t)$
is purely quantum mechanical (first quantization) from 
the region $Y_\infty$ where it is a second quantization of a bosonic field.

\medskip

{\bf 4.6. Oracle assisted QSM system construction.}
It is usually interesting in quantum statistical mechanics to construct 
explicit approximations to systems with infinitely many degrees of 
freedom by systems involving finitely many ones. The  oracle mediated 
construction described in Section 2 provides us with this kind of procedure.  
Consider the sets $A_m$ and $B_m$ 
described in Section 2. We can then consider the algebras
$$ {\Cal A}_m =\otimes_{C\in cp^{-1}(A_m)} {\Cal T}_C, \ \ \ \text{ and } \ \ \
 {\Cal B}_m =\otimes_{C\in cp^{-1}(B_m)} {\Cal T}_C , $$
acting on the tensor product of the Fock spaces, and 
endowed with the tensor product time evolution as above.
Moreover, by denoting, as in the previous section, by $F_m$ the
region between the curves $\Gamma_m$ and $D_m\cap \Gamma_{m-1}$,
one can consider the QSM system associated to the codes with
code points in $F_m$,
$$ {\Cal F}_m =\otimes_{C\in cp^{-1}(F_m)} {\Cal T}_C, \ \ \  \sigma_t= \otimes \sigma_t^C .$$
These give good approximations, by systems involving only finitely many degrees of freedom, 
to the bosonic field theory associated to the set $Y_\infty$ of code parameters and to 
the asymptotic bound $\Gamma$.

\bigskip
\centerline{\bf References}

\medskip

[BaFo] A.~Barg, G.~D.~Forney. {\it Random codes: minimum distances and error exponents.}
IEEE Transactions on Information Theory, vol 48, No. 9 (2002), 2568--2573.

\smallskip

[BoCo] J.~B.~ Bost, A.~ Connes, {\it Hecke algebras, type III factors and phase transitions with spontaneous symmetry breaking in number theory}, Selecta Math. (N.S.) 1 
(1995), no. 3, 411--457.

\smallskip

[BratWe] V.~Brattka, K.~Weihraub. {\it Computability on subsets of Euclidean
space I: closed and compact subsets.} Theoretical Computer Science,
219 (1999), 65--93.

\smallskip

[CoGo] J.T.~Coffey, R.M.~Goodman, {\em Any code of which we cannot think
is good}, IEEE Transactions on Information Theory, Vol.36 (1990) N.6, 1453--1461.

\smallskip

[La]  D.~Lacombe. {\it Extension de la notion de fonction r\'ecursive aux fonctions d'une ou plusieurs
variables r\'eelles., I--III.}\  C.~R.~Ac.~Sci.~Paris, 240 (1955), 2478--2480;
241 (1955), 13--14, 151--153.

\smallskip

[Jul]   B. Julia. {\it Statistical theory of numbers}, in ``Number Theory and Physics, Les
Houches Winter School" (J.-M. Luck, P. Moussa et M. Waldschmidt eds.),
Springer Verlag, 1990.

\smallskip
[Lev1] L.~A.~Levin, {\em Various measures of complexity for finite objects (axiomatic
description)}, Soviet Math. Dokl. Vol.17 (1976) N. 2, 522--526.
\smallskip
[Lev2] L.~A.~Levin, {\em Randomness conservation inequalities; information
and independence in mathematical theories}, Information and Control, Vol. 61 (1984)
15--37.
\smallskip
[LiVi] M.~Li,  P.~M.~B.~Vit\'anyi, {\em An Introduction to Kolmogorov Complexity and its Applications}, 3rd edn. Springer, New York, 2008.

\smallskip

[Man1] Yu.~I.~Manin, {\it What is the maximum number of points
on a curve over $\bold{F}_2$?} J. Fac. Sci. Tokyo, IA, Vol. 28 (1981),
715--720. 

\smallskip

[Man2] Yu.~I.~Manin, {\it  Renormalization and computation I: motivation and background.}
In: Proceedings OPERADS 2009,  J. Loday and B. Vallette eds.,
S\'eminaires et Congr\`es 26, Soc. Math. de France, 2012, pp. 181--223.
Preprint math.QA/0904.4921
\smallskip
[Man3] Yu.~I.~Manin, {\it  Renormalization and Computation II: Time Cut-off and the Halting Problem.}
 In: Math. Struct. in Comp. Science, pp. 1--23, 2012, Cambridge UP.
Preprint math.QA/0908.3430

\smallskip

[Man4] Yu.~I.~Manin, {\em A Course in Mathematical Logic for Mathematicians},
2nd ed., Springer, New York, 2010.
\smallskip
[Man5] Yu.I.~Manin, {\em A computability challenge: asymptotic bounds and 
isolated error-correcting codes},    WTCS 2012 (Calude Festschrift), Ed. by M.J. Dinneen et al.,
 LNCS 7160,  pp. 174--182, 2012. Preprint arXiv:1107.4246.

\smallskip

[ManMar]  Yu.~I.~Manin, M.~Marcolli. {\it Error--correcting codes and phase transitions.}
Mathematics in Computer Science, Vol. 5 (2011), pp. 133--170. arXiv:0910.5135

\smallskip

[ManVla] Yu.~I.~Manin. S.G.~Vladut, {\it Linear codes and
modular curves}.  J. Soviet Math., Vol. 30 (1985),  2611--2643.
\smallskip

[Sch] C.P.~Schnorr, {\em Optimal enumerations and optimal G\"odel
numberings.} Math. Systems Theory, Vol. 8 (1974) N.2, 182--191.

\smallskip

[VlaNoTsfa] S.~G.~Vladut, D.~Yu.~Nogin, M.~A.~Tsfasman. {\it  Algebraic geometric codes: basic notions.} Mathematical Surveys and Monographs, 139. American Mathematical Society, Providence, RI, 2007.

\smallskip

[ZvoLe] A.K.~Zvonkin,  L.A.~Levin, {\em The complexity of finite objects 
and the basing of the concepts of information and randomness on the theory 
of algorithms}, Russ. Math. Surv.  25 (1970) N.6,  83--124.

\enddocument